\documentstyle[11pt]{article}

\title{An elliptic quantum algebra for $\widehat{sl}_2$}

\author{
Omar Foda\thanks{Department of Mathematics, University of Melbourne,
                         Parkville, Victoria 3052, Australia.},
Kenji Iohara\thanks{Department of Mathematics, Faculty of Science,
                            Kyoto University, Kyoto 606, Japan.},
Michio Jimbo$^{\dagger}$,
Rinat Kedem\thanks{Research Institute for Mathematical Sciences,
                            Kyoto University, Kyoto 606, Japan.},\cr
Tetsuji Miwa$^\ddagger$ and
Hong Yan$^\ddagger$\thanks{Address as of April 1994: Institute of
Theoretical Physics, P. O. Box 2735, Beijing, China.} \cr
}

\date{\today}

\begin{document}

\maketitle

\bigskip
\centerline{\sl Dedicated to the memory of Ansgar Schnizer}

\begin{abstract}
An elliptic deformation of $\widehat{sl}_2$ is proposed. Our
presentation of the algebra is based on the relation $RLL=LLR^*$,
where $R$ and $R^*$ are eight-vertex $R$-matrices with the elliptic
moduli chosen differently. In the trigonometric limit, this algebra
reduces to a quotient of that proposed by Reshetikhin and
Semenov-Tian-Shansky.  Conjectures concerning highest weight modules
and vertex operators are formulated, and the physical interpretation
of $R^*$ is discussed.
\end{abstract}
\setcounter{section}{0}

\newfont{\g}{eufm9}
\newcommand{\gtg}{\mbox{\g g}}

\def\C{{\bf C}}
\def\Z{{\bf Z}}
\def\sz{\small {\bf Z}}
\def\H{{\cal H}}
\def\vep{\varepsilon}
\def\z{\zeta}
\def\mod{~{\rm mod}~}
\def\qdet{\hbox{$q$-det}}
\def\id{{\rm id}}
\def\bra#1{\langle #1|}
\def\ket#1{|#1\rangle}
\def\Phim#1{\mathrel{\mathop{\kern0pt \Phi}\limits^#1}}
\def\Psim#1{\mathrel{\mathop{\kern0pt \Psi^*}\limits^#1}}
\def\L#1{\mathrel{\mathop{\kern0pt L}\limits^#1}}
\def\Lm#1{\mathrel{\mathop{\kern0pt L^-}\limits^#1}}
\def\Lp#1{\mathrel{\mathop{\kern0pt L^+}\limits^#1}}
\def\Lpm#1{\mathrel{\mathop{\kern0pt L^\pm}\limits^#1}}
\def\glth{\widehat{gl}_2}
\def\slth{\widehat{sl}_2}
\def\a{{\cal A}}
\def\aqp{{\cal A}_{q,p}(\slth)}
\newtheorem{prop}{Proposition}

\catcode`@=11
\def\citen#1{%
\if@filesw \immediate \write \@auxout {\string \citation {#1}}\fi
\@tempcntb\m@ne \let\@h@ld\relax \def\@citea{}%
\@for \@citeb:=#1\do {%
  \@ifundefined {b@\@citeb}%
    {\@h@ld\@citea\@tempcntb\m@ne{\bf ?}%
    \@warning {Citation `\@citeb ' on page \thepage \space undefined}}%
    {\@tempcnta\@tempcntb \advance\@tempcnta\@ne
    \setbox\z@\hbox\bgroup 
    \ifnum0<0\csname b@\@citeb \endcsname \relax
       \egroup \@tempcntb\number\csname b@\@citeb \endcsname \relax
       \else \egroup \@tempcntb\m@ne \fi
    \ifnum\@tempcnta=\@tempcntb 
       \ifx\@h@ld\relax 
          \edef \@h@ld{\@citea\csname b@\@citeb\endcsname}%
       \else 
          \edef\@h@ld{\hbox{--}\penalty\@highpenalty
            \csname b@\@citeb\endcsname}%
       \fi
    \else   
       \@h@ld\@citea\csname b@\@citeb \endcsname
       \let\@h@ld\relax
    \fi}%
 \def\@citea{,\penalty\@highpenalty}%
}\@h@ld}
\def\@citex[#1]#2{\@cite{\citen{#2}}{#1}}%
\def\@cite#1#2{\leavevmode\unskip
  \ifnum\lastpenalty=\z@\penalty\@highpenalty\fi
  \ [{\multiply\@highpenalty 3 #1
      \if@tempswa,\penalty\@highpenalty\ #2\fi 
    }]\spacefactor\@m}

\catcode`@=12


\section{Introduction}

In this article, we propose an algebra $\a_{q,p}(\slth)$ based on
the $R$-matrix of the eight-vertex model, where $p$ is the
elliptic nome and $q$ is the crossing parameter of the $R$-matrix.

Following the pioneering work of Sklyanin~\cite{Sk1,Sk2}, several
works on elliptic algebras have been published in which finite
dimensional representations were discussed.  For physical applications
it is useful to define an algebra having highest weight modules and
vertex operators. Here, therefore, we define an elliptic algebra and
conjecture the existence of these structures which are natural
deformations of their trigonometric counterparts.  In addition to the
mathematical interest, our motivation in searching for such an algebra
is the wish to understand the structure of the space of states for the
eight-vertex model \cite{8v} in similar way as for the six-vertex and
related models (see \cite{NC} for a review).

One way of introducing a quantum algebra is to use the quantum inverse
scattering method, or `$RLL$ formalism' \cite{FRT,ReshSem}. We follow
a similar path here. In the trigonometric case the algebra is
presented by two generating series, $L^+(\z)$ and $L^-(\z)$, which are
power series in $\z$ and $\z^{-1}$, respectively.  The structure of
the elliptic $R$-matrix, however, forces us to use $L^{\pm}(\z)$ which
are Laurent series in $\z$, involving {\em both} positive and negative
powers of $\z$.  As a result, the number of generators of the algebra
appears to double. We propose to define an elliptic algebra based on a
single relation of the form $RLL=LLR^*$, using a {\em single} $L(\z)$.
Here, the matrices $R,R^*$ are elliptic $R$-matrices, with elliptic
moduli differing by an amount depending on the level $k$ of the
representation on which $L(\z)$ acts.  We show that this definition
resolves the problem of the said doubling.

Since the matrices $R$ and $R^*$ are different, the usual definition
of the coproduct fails.  Nevertheless, the tensor product of two
representations can be defined when one of them has level 0.  Assuming
the existence of highest weight modules, we formulate the vertex
operators as intertwiners, present their commutation relations and
motivate their physical interpretation.  In particular, the physical
$S$-matrix is $S=-R^*$.

Note that in the sine-Gordon model, similar
commutation relations for vertex operators, with $R\neq-S$, were
presented by Lukyanov \cite{luk1,luk2}.

The text is organized as follows.  We fix the notations for the
$R$-matrix in section 2.  In section 3 the elliptic algebra is
defined. The reduction to the trigonometric case is discussed in
section 4.  In section 5 we formulate conjectures regarding the
highest weight modules and vertex operators, in the level one case.
Section 6 contains a discussion and remarks.

\section{The $R$-matrices}

The elliptic $R$-matrix of the eight-vertex model is a function of
three parameters, $p,~q$ and $\z$. In terms of the parameters
$I,~I',~\lambda$ and $u$ introduced in~\cite{Baxbk} (cf. (10.4.23--24) and
(10.7.9)),
\begin{equation}
p=\exp\left(-\frac{\pi I'}{I}\right),
\quad q=-\exp\left(-\frac{\pi\lambda}{2I}\right),
\quad \zeta=\exp\left(\frac{\pi u}{2I}\right).
\label{baxparam}
\end{equation}

We introduce two $R$ matrices,
\[
R^\pm(\z)=R^\pm(\zeta;p^{1/2},q^{1/2}),
\]
which have the structure
\begin{equation}\label{Rpm}
R^\pm(\zeta)=
\left(
\begin{array}{cccc}
a^\pm(\zeta)&&&d^\pm(\zeta)\\
&b^\pm(\zeta)&c^\pm(\zeta)&\\
&c^\pm(\zeta)&b^\pm(\zeta)&\\
d^\pm(\zeta)&&&a^\pm(\zeta)\end{array}\right).
\end{equation}
When viewed as holomorphic functions of the complex variables
$p^{1/2},~q^{1/2}$ and $\z$ with $|p^{1/2}|<|q|<1$, $R^+$ and $R^-$
coincide with the $R$-matrix of \cite{Baxbk} up to scalar
factors. Here we adopt the formal series point of view, regarding
$p^{1/2}$, $q^{1/2}$ and $\z$ as indeterminates, and $R^\pm$ as a formal
series in $\z$.

The entries of (\ref{Rpm}) are given as follows.
Let $B=\C(q^{1/2})[[p^{1/2}]]$ denote the ring of formal power series
in $p^{1/2}$ whose coefficients are rational functions in $q^{1/2}$.
Introduce the formal power series with coefficients in $B$
\begin{eqnarray}
\rho(\zeta^2)&=&q^{-1/2}\displaystyle
\frac{(q^2\zeta^2;q^4)_\infty^2}
{(\zeta^2;q^4)_\infty(q^4\zeta^2;q^4)_\infty},
\label{rho}\\
\bar{\alpha}(\zeta)&=&\displaystyle
\frac{(p^{1/2}q\zeta;p)_\infty}
{(p^{1/2}q^{-1}\zeta;p)_\infty}
\displaystyle \frac{(pq^4\zeta^2;p,q^4)_\infty(p\zeta^2;p,q^4)_\infty}
{(pq^2\zeta^2;p,q^4)^2_\infty},
\label{alpha}\\
\displaystyle \bar{\beta}(\zeta)&=&\displaystyle
\frac{(-pq\zeta;p)_\infty}{(-pq^{-1}\zeta;p)_\infty}
\displaystyle \frac{(pq^4\zeta^2;p,q^4)_\infty(p\zeta^2;p,q^4)_\infty}{
(pq^2\zeta^2;p,q^4)^2_\infty},
\label{beta}
\end{eqnarray}
where
\begin{eqnarray}
(z;p_1,\cdots,p_m)_\infty
&=&\prod_{n_1,\cdots,n_m\ge 0}(1-zp_1^{n_1}\cdots p_m^{n_m})\nonumber\\
&=&\exp\left(-\sum_{k=1}^\infty\frac{1}{(1-p_1^k)\cdots(1-p_m^k)}
\frac{z^k}{k}\right).\nonumber
\end{eqnarray}
To specify the entries of (\ref{Rpm}), we demand that
$a^\pm(\z),~b^\pm(\z)$ be even in $\z$, and $c^\pm(\z),~d^\pm(\z)$ be
odd. We set
\begin{equation}
\label{R-entries}
\begin{array}{rcl}
a^\pm(\zeta)+d^\pm(\zeta)&=&\rho(\zeta^{\pm2})^{\pm 1}
\displaystyle
\frac{\bar{\alpha}(\zeta^{-1})}{\bar{\alpha}(\zeta)}~,\\
b^\pm(\zeta)+c^\pm(\zeta)&=&\rho(\zeta^{\pm 2})^{\pm 1}
q^{\pm 1}
\displaystyle
\frac{1+(q^{-1}\zeta)^{\pm 1}}{1+(q\zeta)^{\pm 1}}
\displaystyle
\frac{\bar{\beta}(\zeta^{-1})}{\bar{\beta}(\zeta)}~.
\end{array}
\end{equation}
In the second equation the factor $(1+(q\z)^{\pm1})^{-1}$ should be
expanded in powers of $(q\z)^{\pm 1}$.

One can verify that the coefficients of the series
$R^\pm(\z)={\displaystyle \sum_{n\in \sz}} R^\pm_n \z^n$ satisfy
\[
R^\pm_n\equiv 0~~ \mod  \bigl(p^{1/2}\bigr)^{\max\,(\mp n,0)}\!B
\qquad \forall n\in\Z.
\]
Note that when $p=0$,  $R^+(\z)$ (resp. $R^-(\z)$) contains only non-negative
(resp. non-positive) powers in $\z$.

We regard $R^\pm(\z)$ as linear operators on $V\otimes V$, with
$V=Bv_+\oplus Bv_-$, and set $R^\pm(\z)v_{\vep'_1}\otimes v_{\vep'_2}
=\sum v_{\vep_1}\otimes v_{\vep_2}
R^\pm(\z)_{\vep_1\vep_2;\vep_1'\vep_2'}$.  When written in the matrix
form (\ref{Rpm}), the entries of $R^\pm$ are arranged in the order
$(\vep_1,\vep_2)=(++),(+-),(-+),(--)$. The $R$-matrices have the
following properties:
\begin{description}
\item[Yang-Baxter equation]
\begin{equation}\label{YBE}
R^\pm_{12}(\z_1/\z_2)R^\pm_{13}(\z_1/\z_3)R^\pm_{23}(\z_2/\z_3)
=
R^\pm_{23}(\z_2/\z_3)R^\pm_{13}(\z_1/\z_3)R^\pm_{12}(\z_1/\z_2),
\end{equation}
\item[Unitarity]
\begin{equation}\label{unitarity}
R^\pm_{12}(\z_1/\z_2)R^\mp_{21}(\z_2/\z_1)=\id,
\end{equation}
\item[Crossing symmetry]
\begin{equation}\label{crossing}
R^\pm_{21}(\z_2/\z_1)^{t_1}=
\sigma^x_1 R^\mp_{12}(-q^{-1}\z_1/\z_2)\sigma^x_1,
\end{equation}
\item[Quasi-periodicity]
\begin{equation}\label{pro1}
R^\pm_{12}(-\zeta)=\sigma_1^zR^\pm_{12}(\zeta)\sigma_1^z=
\sigma^z_2R^\pm_{12}(\zeta)\sigma^z_2~,
\end{equation}
\begin{equation}\label{pro2}
R^+_{12}(-p^{1/2}\zeta)=\sigma_1^xR^-_{12}(\zeta)\sigma_1^x=
\sigma^x_2R^-_{12}(\zeta)\sigma^x_2~.
\end{equation}
\end{description}
Here, if $R^\pm(\z)=\sum a_i\otimes b_i$, with $a_i,~b_i
\in {\rm End}(V)$, then
$R^\pm_{21}(\z)=\sum b_i\otimes a_i$,
$R^\pm_{13}(\z)=\sum a_i\otimes \id \otimes b_i$, etc..
The Pauli matrices are chosen to be
\[
\sigma^x=\left(\matrix{0&1\cr 1&0\cr}\right),
\qquad
\sigma^y=\left(\matrix{0&-i\cr i&0\cr}\right),
\qquad
\sigma^z=\left(\matrix{1&0\cr 0&-1\cr}\right),
\]
the suffix $j(=1,2)$ indicating that they are acting on the $j^{\rm
th}$-component.

The formulas (\ref{pro1})--(\ref{pro2}) are the formal series version of
the characterizing properties of the elliptic
$R$-matrix in Belavin's approach \cite{Bel}.

\section{The algebra {$\aqp$}}

Let us proceed to the definition of the elliptic algebra.
Consider a free associative algebra ${\cal U}^{''}$
over $\C(q^{1/2})$ on the letters
$\bar{L}_{\vep\vep',n}$,
where $n\in$ {\bf Z}, $\vep,\vep'=\pm$ and
$\vep\vep'=(-1)^n$.
By convention we set
\begin{equation}\label{ParL}
\bar{L}_{\vep\vep',n}=0
\qquad \hbox{ if }\vep\vep'\neq (-1)^n.
\end{equation}
Let ${\cal U}^{\prime}$ denote the algebra obtained by adjoining an
invertible central element $q^{c/2}$ to ${\cal
U}^{''}\otimes_{\C(q^{1/2})} B$. Let further ${\cal U}={\displaystyle
{\lim_\leftarrow}}~ {\cal U}'/p^{N/2}{\cal U}'$ be the $p^{1/2}$-adic
completion of ${\cal U}^\prime$.

Consider the formal series with entries in ${\cal U}$
\begin{equation}
L(\zeta)=\left(
\begin{array}{cc}
L_{++}(\zeta)&L_{+-}(\zeta)\\
L_{-+}(\zeta)&L_{--}(\zeta)
\end{array}\right)
\end{equation}
where
\[
L_{\vep\vep'}(\zeta)=\sum_{n\in\sz} L_{\vep\vep',n}\z^{-n},
\qquad
L_{\vep\vep',n}=(-p^{1/2})^{\max\,(n,0)}\bar{L}_{\vep\vep',n}.
\]

We define the algebra ${\cal A}_{q,p}(\glth)$ by imposing the following
relations on $L(\z)$:
\begin{equation}\label{RLL}
R^+_{12}(\zeta_1/\zeta_2)\L{1}(\zeta_1)\L{2}(\zeta_2)
=\L{2}(\zeta_2)\L{1}(\zeta_1)R^{*+}_{12}(\zeta_1/\zeta_2)~,
\end{equation}
where
\[
\L{1}(\zeta)= L(\z)\otimes  {\rm id},\qquad
\L{2}(\zeta)=  {\rm id} \otimes L(\z),
\]
and
\begin{equation}
R^{*+}(\z)=R^+(\z;p^{*1/2},q^{1/2}),
\qquad p^{*1/2}=p^{1/2}q^{-c}.
\end{equation}
To be precise, consider the difference (LHS)$-$(RHS) of (\ref{RLL})
and take its coefficients for various powers of $\z_1,\z_2$.
Their matrix entries are well defined elements of ${\cal U}$.
Let ${\cal I}$ denote the ideal generated by these matrix elements,
and let $\overline{\cal I}$ denote its closure in the
$p^{1/2}$-adic topology.
Then we define ${\cal A}_{q,p}(\glth)={\cal U}/\overline{\cal I}$.

By a standard argument
based on equations (\ref{YBE})--(\ref{crossing})
we find that
the following quantum determinant
belongs to the center of ${\cal A}_{q,p}(\glth)$:
\[
\qdet L(\z)= L_{++}(\z/q)L_{--}(\z)-L_{-+}(\z/q)L_{+-}(\z).
\]
Imposing further the relation
$\qdet L(\z)= q^{c/2}$ we define the quotient algebra
\[
\aqp={\cal A}_{q,p}(\glth)/\langle \qdet L(\z)- q^{c/2}\rangle.
\]

Unlike in the trigonometric case one cannot define the notion of
weights since the $R$-matrix does not have a spin-conservation
property, i.e.
does not commute with matrices of the form $h\otimes h$ with
$h$ an arbitrary diagonal matrix.
Nevertheless the algebra ${\cal A}_{q,p}(\glth)$ admits a $\Z$-grading
\[
\deg L_{\vep\vep',n}=-n
\]
which corresponds to the principal grading for
affine Lie algebras.
Clearly $\aqp$ inherits this grading as well.

\section{Reduction to the trigonometric case}
The formulation of \cite{ReshSem} involves {\em two} $L$-operators,
$L^\pm(\z)$. At $p=0$, our algebra $\a_{q,p}(\glth)$ reduces to a
factor algebra of $A(R)$ of \cite{ReshSem}, in the following sense.
The algebra $A(R)$, with the conditions $L^\pm_n = 0$ if $\mp n > 0$
and $\tilde{L}^\pm(\z)=(L^\pm(\z)^t)^{-1}$ (see eq. 3.21 of \cite{DF})
imposed, is our $\a_q(\glth)\stackrel{\rm def}{=} \a_{q,0}(\glth)$. To
see this, define the auxiliary $L$-operators
\[
L^+(\z)=L(q^{c/2}\z),
\qquad
L^-(\z)=\sigma^x L(-p^{1/2}\z) \sigma^x.
\]
These are Laurent series of the form
\begin{eqnarray*}
L^+(\z)&=&
\cdots + p^{1/2}L^+_1 \z^{-1} + L^+_0 \z^0 + L^+_{-1} \z +\cdots,
\\
L^-(\z)&=&
\cdots + L^-_1 \z^{-1} + L^-_0 \z^0 + p^{1/2}L^-_{-1} \z +\cdots,
\end{eqnarray*}
and hence at $p=0$ they become
power series in $\z^{\pm 1}$, respectively.
Using (\ref{pro2}) one can show

\begin{prop}
The following relations hold in ${\cal A}_{q,p}(\glth)$.
\begin{eqnarray*}
&&R^\pm_{12}(\zeta_1/\zeta_2)\Lpm{1}(\zeta_1)\Lpm{2}(\zeta_2)
=\Lpm{2}(\zeta_2)\Lpm{1}(\zeta_1)R^{*\pm}_{12}(\zeta_1/\zeta_2),
\\
&&R^+_{12}(q^{c/2}\zeta_1/\zeta_2)\Lp{1}(\zeta_1)\Lm{2}(\zeta_2)
=\Lm{2}(\zeta_2)\Lp{1}(\zeta_1)R^{*+}_{12}(q^{-c/2}\zeta_1/\zeta_2).
\end{eqnarray*}
\end{prop}
These are the defining relations used in \cite{ReshSem}. We remark that
it is due to the scaled nome in $R^*$ that these three equations follow
from the single equation (\ref{RLL}).

Note that there exist three different presentations of the quantum affine
algebra $U_q(\slth)$. The first two, using the Chevalley generators
\cite{Drin1,Jimbo1} and using the Drinfeld generators
\cite{drinfeld}, were shown by Beck \cite{Beck} to be isomorphic.
A third, isomorphic, presentation is that of Ding and Frenkel
\cite{DF}.  The algebra $\a_q(\slth)$ is similar to that of
\cite{DF}, the $R$-matrices differing by a scalar factor. It
appears that the precise relation between $\a_q(\slth)$ and
$U_q(\slth)$ remains an open question.

\section{Vertex operators}
The simplest representation of $U=\aqp$ is the analog
of the spin-$1/2$ evaluation module
of ${\cal A}_{q}(\slth)$.
Let $V_\xi=V\otimes B[\xi,\xi^{-1}]$.
Then $V_\xi$ becomes a $U$-module with the assignment
\[
L_{\vep_1\vep_1'}(\z)v_{\vep_2'}=
\sum_{\vep_2} v_{\vep_2}R^+(\z/\xi)_{\vep_1\vep_2;\vep_1'\vep_2'},
\qquad q^{c/2}=1.
\]
In general we say that a $U$-module has level $k\in \Z$
if the central element $q^{c/2}$ acts as a scalar $q^{k/2}$.
Thus $V_\xi$ has level $0$.

For physical applications
we are interested in representations with non-zero level.
In what follows we will formulate conjectures concerning highest
weight representations of level $1$.

We expect the following to hold:
\begin{description}
\item[Highest weight modules]
There exist $U$-modules $\H^{(i)}$ ($i=0,1$) generated by a single
vector $\ket{i}$ with the properties
\begin{eqnarray*}
&&L_{\vep\vep',n}\ket{i}=0~~(n>0),
\quad  L_{++,0}\ket{i}=a\, q^{(1-i)/2}\ket{i},
\quad  L_{--,0}\ket{i}=a\, q^{i/2}\ket{i}
\end{eqnarray*}
where $a$ is a scalar to be specified below (see (\ref{gamma})).
They are $\Z$-graded:
$\H^{(i)}=\oplus_{d=0}^\infty \H^{(i)}_d$, and have the
character
\begin{equation}\label{ch}
\sum_{d=0}^\infty {\rm dim\,}\H^{(i)}_d~t^d~
=\prod_{j=1}^\infty (1+t^j).
\end{equation}
\item[Vertex operators]
There exist intertwiners of $U$-modules
\begin{eqnarray*}
\Phi(\xi)&:& \H^{(i)} \longrightarrow \H^{(1-i)}\otimes V_\xi,
\qquad \Phi(\xi)=\sum\Phi_\vep(\xi)\otimes v_\vep,
\\
\Psi^*(\xi)&:& V_\xi\otimes \H^{(i)} \longrightarrow \H^{(1-i)},
\qquad \Psi^*_\vep(\xi)= \Psi^*(\xi)\left(v_\vep\otimes\cdot\right),
\end{eqnarray*}
which we call vertex operators of type I and type II, respectively.
To be more precise, the vertex operators are formal series
\begin{eqnarray*}
\Phi_\vep(\xi)&=&\sum_{n \equiv i+(1+\vep)/2 \mod 2} \Phi_{\vep,n}\xi^{-n},
\\
\Psi^*_\vep(\xi)&=&\sum_{n \equiv i+(1-\vep)/2\mod 2} \Psi^*_{\vep,n}\xi^{-n},
\end{eqnarray*}
with components
$\Phi_n,\Psi^*_n:{\cal H}^{(i)}_d \rightarrow {\cal H}^{(1-i)}_{d-n}$,
satisfying the intertwining relations
\begin{eqnarray}
\Phi_{\vep_2}(\z_2) L_{\vep_1,\vep_1''}(\z_1)
&=&
\sum
R^+_{\vep_1 \vep_2;\vep_1'\vep_2'}(\z_1/\z_2)
L_{\vep_1'\vep_1''}(\z_1)\Phi_{\vep_2'}(\z_2),
\label{LPhi}\\
L_{\vep_2\vep_2''}(\z_2)\Psi^*_{\vep_1''}(\z_1)
&=& \sum
\Psi^*_{\vep_1'}(\z_1)L_{\vep_2\vep_2'}(\z_2)
R^{*+}_{\vep_2'\vep_1';\vep_2''\vep_1''}(q^{-1/2}\z_2/\z_1).\nonumber\\
\label{LPsi}
\end{eqnarray}
We normalize the vertex operators by setting
$\Phi_0\ket{i}=\ket{1-i}$, $\Psi^*_0\ket{i}=\ket{1-i}$.

\item[Commutation relations]
The vertex operators satisfy the commutation relations
\begin{eqnarray}
\sum R_{\vep_1\vep_2;\vep_1'\vep_2'}(\z_1/\z_2)
\Phi_{\vep_1'}(\z_1)\Phi_{\vep_2'}(\z_2)
&=&
\Phi_{\vep_2}(\z_2)\Phi_{\vep_1}(\z_1),
\label{PhiPhi}
\\
\tau(\z_1/\z_2)\Psi^*_{\vep_2}(\z_2)\Phi_{\vep_1}(\z_1)
&=&
\Phi_{\vep_1}(\z_1)\Psi^*_{\vep_2}(\z_2),
\label{PhiPsi}
\\
-\sum \Psi^*_{\vep_2}(\z_2)\Psi^*_{\vep_1}(\z_1)
R^*_{\vep_1\vep_2;\vep_1'\vep_2'}(\z_1/\z_2)
&=&
\Psi^*_{\vep_1'}(\z_1)\Psi^*_{\vep_2'}(\z_2).
\label{PsiPsi}
\end{eqnarray}
Here
\begin{eqnarray}
&&\tau(\z)=\z^{-1}
\frac{(q\z^2;q^4)_\infty}{(q^3\z^2;q^4)_\infty}
\frac{(q^3\z^{-2};q^4)_\infty}{(q\z^{-2};q^4)_\infty},
\label{tau}\\
&&R(\z)=\tau(q^{1/2}\z^{-1})^{-1} R^+(\z).
\nonumber
\end{eqnarray}
Further, the inversion property
\begin{equation}\label{inv}
g \sum_\vep
\Phi_{-\vep}(-\zeta/q)\Phi_{\vep}(\zeta)={\rm id}
\end{equation}
holds for type I operators where the scalar $g$ is defined in (\ref{g})
below.
\end{description}

These conjectures are direct generalizations of the known
structures for the trigonometric case \cite{FreResh,DFJMN}.
We expect in particular the character (\ref{ch}) to remain the same.
The existence of such spaces $\H^{(i)}$ and
vertex operators have been proposed earlier in \cite{8v,Ising},
by physical arguments based on the corner transfer matrix method.
The elliptic algebra and the type II operators were not
discussed there. Commutation relations of the form
(\ref{PhiPhi})--(\ref{PsiPsi}), with $R\neq R^*$, appeared in the
sine-Gordon case in \cite{luk1}, equation (7.2).

One can rewrite the commutation relations (\ref{RLL}) for the components
as a `normal ordering rule' of the form
\[
L_{\vep_1\vep_1',m}L_{\vep_2\vep_2',n}
=
\sum_{\sigma_1,\cdots,\sigma_2'}\sum_{j\ge 0}
C_j L_{\sigma_2\sigma_2',(m+n+s)/2-j}L_{\sigma_1\sigma_1',(m+n-s)/2+j}
\]
where $s=0,1$ according to whether $m\equiv n\mod 2$ or not.  The
coefficients $C_j$ (depending on all $\vep_1,\cdots,\sigma_1,\cdots$,
$m,n$) are written in terms of the Taylor coefficients of  known functions
such as $\bar{\alpha}(\z)$ and $\bar{\beta}(\z)$. This makes it
possible in principle to calculate matrix elements for products of
$L$-operators.
For instance we have
\[
\bra{i}\Bigl(L_{++}(\z_1)L_{--}(\z_2)+L_{-+}(\z_1)L_{+-}(\z_2)\Bigr)\ket{i}
=
a^2\,q^{1/2}\,\frac{\bar{\beta}^*(\z_2/\z_1)}{\bar{\beta}(\z_2/\z_1)}
\]
where $\bar{\beta}^*(\z)$ denotes $\bar{\beta}(\z)$ in
(\ref{beta}) with $p$ replaced by $p^*=p q^{-2}$.
We find that in order to satisfy the quantum determinant relation
one must choose $a$ to be
\begin{equation}\label{gamma}
a^2=\frac{\bar{\beta}(-q)}{\bar{\beta}^*(-q)}.
\end{equation}
In the same way one can calculate the matrix elements of products of
$\Phi$ operators to obtain
\begin{equation}\label{g}
g=\bar{\beta}(-q)g_0,
\qquad
g_0=\frac{(q^2;q^4)_\infty}{(q^4;q^4)_\infty}.
\end{equation}

In fact the type II operator in the trigonometric case can be expressed as
\begin{equation}\label{PsiLPhi}
\Psi^*_{\vep'}(\z)=a^{-1}\bar{\beta}(-q)
\sum_{\vep} L_{\vep\vep'}(q^{1/2}\z)\Phi_{-\vep}(-q^{3/2}\z),
\end{equation}
since both sides have the same intertwining property.
Using the same definition in the elliptic case, the commutation relations
(\ref{LPsi}), (\ref{PhiPsi}), (\ref{PsiPsi}) involving $\Psi^*$
are consequences of
(\ref{RLL}), (\ref{LPhi}) and (\ref{PhiPhi}).
We remark that conversely the inversion property (\ref{inv})
enables us to write $L(\z)$ as
\[
L_{\vep\vep'}(\z)=a g_0 \Psi^*_{\vep'}(q^{-1/2}\z)\Phi_\vep(\z).
\]
This formula has been pointed out in the trigonometric case
by Miki \cite{Miki}.

The highest weight modules for generic level can be formulated in the
same way, and proper generalizations of the relations in this section hold.
Similarly, it is straightforward to extend the present formulation to
the case of $sl_n$.

\section{Discussion}

As discussed in \cite{8v}, type I vertex operators correspond to the
half-column transfer matrix of the eight-vertex model. On the other
hand the type II operators create the
eigenstates
of the transfer matrix, the factor $\tau(\z)$ being the corresponding
single-particle
eigenvalue.  Such an interpretation was established in the trigonometric
case in
\cite{DFJMN} (see also \cite{NC}). It is known \cite{JKM} that the
single-particle excitation energy for the eight-vertex model does not
depend on the elliptic modulus, and therefore is the same as in the
trigonometric case. Our commutation relation (\ref{PhiPsi}) agrees
with that picture.

In the trigonometric case the $R$-matrix entering the $\Psi^*\Psi^*$
commutation relation is interpreted as the $S$-matrix for the
excitations.  For the eight-vertex model, we could not find the
calculation of the $S$-matrix in the literature.  However, it is known
that in the limit $I\rightarrow\infty$ with $\lambda,~u$ fixed, in
terms of the parameters of (\ref{baxparam}), the
eight-vertex model scales to massive Thirring/sine-Gordon field
theory. Considering the $S$-matrix of this theory, Smirnov has
suggested to us that the $S$-matrix of the eight-vertex model is given
by $-R^*(\z)$ in the region $0<p\leq q^4$, where there are no bound
states~\cite{JKM}.

There is another special case worth mentioning. As is well known, at
$p=q^4$ the eight-vertex model decouples to two non-interacting Ising
models. An independent construction of $\H^{(i)}$ and $\Phi(\z)$ was
given in~\cite{Ising} using free fermions. In this case
$-R^*(\z)=-\sigma^y\otimes\sigma^y$.  Passing to the basis in which
$\sigma^y$ is diagonal, we see that there are two species of
excitations, those of the same kind mutually anti-commute and those of
different kind commute.  This agrees with the well known fact that for
the single Ising model the $S$-matrix is $-1$.

There are several questions which arise:
Is it possible to
find an analog of the Chevalley generators which satisfy relations
involving a finite number of terms, as do Uglov's generators \cite{Ug}
for level $0$?
What is the relation between our algebra and the Sklyanin algebra,
or Baxter's intertwining vectors \cite{Baxter,SOS}?
How should one understand the lack of a Hopf algebra structure?
Is there an analog of the universal $R$-matrix?
Do the matrix elements of products of vertex operators satisfy
difference equations analogous to the q-KZ equations~\cite{FreResh}?
Can one find an integral formula for them by
bosonizing the level one modules?

These remain interesting open problems.

\vskip 1cm
\noindent{\sl Acknowledgements.\quad}
We would like to thank F.~Smirnov and E.~Sklyanin for their interest
and enlightening discussions. We also thank S.~Lukyanov for drawing
our attention to his work.  This work is partly supported by
Grant-in-Aid for Scientific Research on Priority Areas 231, the
Ministry of Education, Science and Culture.  R.K. and H.Y. are
supported by the Japan Society for the Promotion of Science. O.F. is
supported by the Australian Research Council. H.Y. would like to thank
M.~Kashiwara for frequent discussions and constant encouragement and
RIMS for warm hospitality.

\bibliographystyle{unsrt}

\end{document}